\documentclass[
 reprint,
 amsmath,
 pre,
 amssymb,
 nofootinbib,
 aps, titlepage
]{revtex4}  %{revtex4-1}  M.K.: 4-1 does not work for me, we can revert later. 

\usepackage[pdftex]{graphicx}
\usepackage{wasysym}
\usepackage{amsfonts}
\usepackage{ifsym}

\makeatletter
\newlength{\apb@width}
\newcommand{\autoparbox}[2][c]{\settowidth{\apb@width}{#2}\parbox[#1]{\apb@width}{#2}}

\makeatother

\newcommand{\namedref}[2]{\hyperref[#2]{#1~\ref*{#2}}}

\newcommand{\vev}[1]{\langle#1\rangle}

\newcommand{\Csphere}{{}^\bullet\kern-1.2pt C}
\newcommand{\Ctorus}{{}^\circ\kern-1.2pt C}

\newcommand{\COMMENT}[1]{}

\newcommand{\neqa}{\nonumber\end{eqnarray}}

\newcommand{\<}{{\langle}}
\renewcommand{\>}{{\rangle}}

\newcommand{\re}{\relax{\rm I\kern-.18em R}}

\def\su2{{SU(2)}}

\def\[{\left[}
\def\]{\right]}

\def\({\left(}
\def\){\right)}
\def\[{\left[}
\def\]{\right]}

\def\<{\langle}
\def\>{\rangle}

\def\i2{\frac{i}{2}}

\def\2F1{\,_2{\rm F}_1}

\usepackage[usenames,dvipsnames]{xcolor}
\usepackage{color}
\usepackage{graphicx}
\usepackage{dcolumn}
\usepackage{bm}
\usepackage{hyperref}

\usepackage{cancel}
\usepackage{ctable}
\usepackage{booktabs}
\newcolumntype{L}[1]{>{\raggedright\let\newline\\\arraybackslash\hspace{0pt}}m{#1}}
\newcolumntype{C}[1]{>{\centering\let\newline\\\arraybackslash\hspace{0pt}}m{#1}}
\newcolumntype{R}[1]{>{\raggedleft\let\newline\\\arraybackslash\hspace{0pt}}m{#1}}

\newcommand{\beq}{\begin{equation}}
\newcommand{\eeq}{\end{equation}}
\newcommand{\beqq}{\begin{equation*}}
\newcommand{\eeqq}{\end{equation*}}
\newcommand\beqa{\begin{eqnarray}}
\newcommand\eeqa{\end{eqnarray}}
\newcommand\beqaa{\begin{eqnarray*}}
\newcommand\eeqaa{\end{eqnarray*}}
\newcommand\bea{\begin{array}}
\newcommand\eea{\end{array}}

\begin{document}

%\preprint{APS/123-QED}

 \begin{titlepage}

\title{Snowmass White Paper:  S-matrix Bootstrap}

\author{Martin Kruczenski}
\affiliation{Department of Physics and Astronomy \\
Purdue University, West Lafayette, IN 47907, USA.}
\author{Jo\~ao Penedones}
\affiliation{Fields and String Laboratory, Institute of Physics, \'Ecole Polytechnique F\'ed\'erale de Lausanne (EPFL), \\
Rte de la Sorge, BSP 728, CH-1015 Lausanne, Switzerland}
\author{Balt C. van Rees}
\affiliation{CPHT, CNRS, Ecole Polytechnique, Institut Polytechnique de Paris, \\ Route de Saclay, 91128 Palaiseau, France
\vskip 2cm
\includegraphics[width=0.3\textwidth]{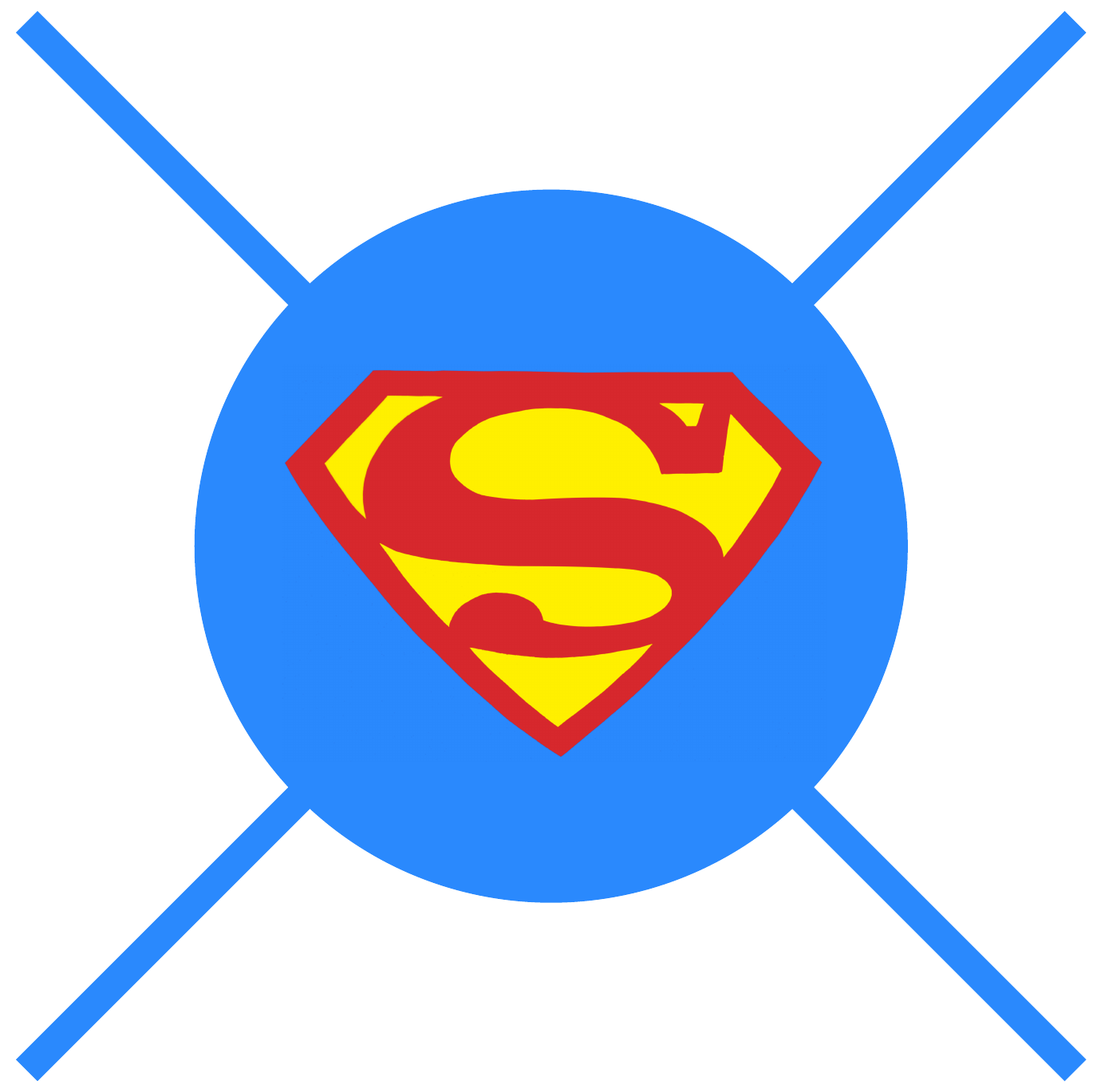}
\vskip 2cm
} 


\begin{abstract}
The S-matrix Bootstrap originated on the idea that the S-matrix might be fully constrained by global symmetries, crossing, unitarity, and analyticity without relying on an underlying dynamical theory that may or may not be a quantum field theory. Recently this approach was revived from a somewhat different point of view. Using the same constraints, one numerically maps out the (infinite-dimensional) space of allowed S-matrices that should contain all consistent quantum field theories (and quantum theories of gravity). Moreover, in the best case scenario one finds special points in the space that can be identified with a certain quantum field theory of interest. In that case, the approach allows the numerical computation of the S-matrix without relying on the particular Lagrangian of the theory. %As the old approach, the new one is most easily applied in two dimensiogons but given the vast improvements in numerical power and techniques it proves its worth when applied to 3+1 dimensions.

In this white paper we summarize the state of the art and discuss the future of the topic.
\end{abstract}

\pacs{Valid PACS appear here}% PACS, the Physics and Astronomy

\maketitle

\end{titlepage}

% \newpage

\section{Introduction}
\label{sec:introduction}
In the middle of the twentieth century physicists were struggling to understand both the forces that hold together the nuclear constituents as well as the numerous newly discovered (mostly unstable) hadronic particles. A popular idea at the time was nuclear democracy: all particles are on an equal footing and none is more elementary than others. Furthermore, all these particles must have scattering amplitudes compatible with the principles of Lorentz invariance, causality and unitarity. This led to the old S-matrix Bootstrap: the idea that imposing such consistency conditions would fully determine the theory.

In hindsight this idea was far too optimistic. Indeed, given the infinity of presumably consistent QFTs each of these has an S-matrix compatible with all the bootstrap principles. It is more realistic to quantitatively constrain the space of consistent QFTs non-perturbatively. This is essentially the aim of the modern S-matrix bootstrap program. 

In practice numerical algorithms are often used to implement the principles in a pragmatic and efficient way. In this way the community has begun to explore the landscape of scattering amplitudes, and uncovered universal bounds and extremal amplitudes in different settings. In the following we will provide a summary of the S-matrix bootstrap methods and survey the results obtained so far. We will sketch a brief vision for the future in the final two sections.

\section{State of the art}
\label{sec:sota}

\subsection{Analyticity}
There is no S-matrix bootstrap without analyticity. Nevertheless we only have an incomplete picture of the analyticity properties of scattering amplitudes. What we do know relies mostly on the LSZ prescriptions together with some auxiliary assumptions like the distributional properties of position-space correlation functions. Starting from these, early results on the analyticity of two-to-two scattering amplitudes in gapped theories include the small and large Lehmann ellipses as well as analyticity in $s$ for fixed $t \leq 0$ in specific cases. A big step forward was made by Bros, Epstein and Glaser \cite{Bros:1965kbd} who, with considerable effort, proved \emph{crossing symmetry}: for two-to-two scattering, the $s$-, $t$-, and $u$-channel amplitudes are the boundary values of a single analytic function. Martin \cite{Martin:1965jj} then combined their results with the constraints of unitarity, which allowed him in particular to show that the analyticity in $s$ persists for positive $t < t^*$, with $t^* > 0$ determined by the physics in the crossed channel.  This was instrumental for his rigorous proof of the Froissart(-Martin) bound on the total cross section. Unfortunately very little progress has been made in the half-century that has since passed, and the question of analyticity remains wide open for general values of the Mandelstam invariants, on other sheets, or for higher-point amplitudes. For the interested reader we remark that old reviews of these results include \cite{Martin:1965jj,Sommer:1970mr}; a more modern comprehensive discussion is contained in the textbook \cite{Bogolyubov:1990kw}. See also the recent paper \cite{Correia:2020xtr} for an efficient summary. 

Of course, the optimistic physicist is not deterred by this dearth of rigorous results. It is believed that the domain of analyticity of any scattering amplitude, and in particular of two-to-two scattering, is significantly larger than what is rigorously proven. One particular take is to assume \emph{Landau} analyticity. This has come to mean the idea that all the singularities in the scattering amplitude (on the principal sheets) have a Landau diagram interpretation. The easiest process from an analyticity viewpoint is two-to-two scattering of the lightest particle in the theory. Then Landau analyticity should imply a fully analytic amplitude outside the cuts at $s, t, u \geq 4m^2$ and any potential bound-state poles. More generally Landau analyticity would hold, essentially by definition, if the sum over Feynman diagrams would converge. The S-matrices for two-dimensional integrable theories also appear to be in agreement with Landau analyticity, although it can take some work to show this \cite{Coleman:1978kk}. Some very recent analyses of Landau singularities can be found in \cite{Mizera:2021ujs,Mizera:2021fap,Correia:2021etg}. 

There is an intriguing complementarity between the axiomatic results and the Landau diagram analysis. On the one hand Landau diagrams inform us about specific non-analyticities like anomalous thresholds. They also have a very physical interpretation as particles travelling over large distances and interacting at widely separated points. But by construction these singularities must occur precisely outside the domain where the axiomatic analytic results apply. For general amplitudes there remains quite literally a gap between these approaches. It would be fascinating to see to which extent this can be bridged in the future.

\subsection{Two-dimensional amplitudes  }
\label{sec:2D}
In the introduction we stated that it would be too optimistic to think that any non-trivial S-matrix could be fully determined by general consistency conditions. There is one exception to this rule, which are the integrable field theories. Their existence is due to the simplicity of two-dimensional scattering, and for the same reason $d=2$ provided a convenient setting for the revived S-matrix bootstrap. 

A first recent work in this direction was \cite{Paulos:2016but}, who rediscovered an old result \cite{Creutz:1973rw} and made contact with QFT in AdS discussed below. The main idea advocated in that paper was to define a basis for the space of all amplitudes allowed by %considerations 
analyticity and crossing symmetry, and then impose the non-linear unitarity condition afterwards, often numerically. (We will review the higher-dimensional version of this approach in the next subsection.) The specific example of \cite{Paulos:2016but} concerned two-to-two scattering for a single species of particles with mass $m$ with exactly one bound state below threshold, corresponding to a pole in the amplitude on the physical sheet. Physically one expects the coupling of the particle to the bound state, so the residue of the pole, to be bounded because new bound states ought to appear if it were very strongly coupled. Such an upper bound indeed exists, and in the above works it was computed both numerically and analytically.

This initial result was quickly generalized. In \cite{He:2018uxa} it was observed that even with no bound states the space of two-to-two amplitudes was restricted and, in fact, it was shown that the space is convex. In the case of flavor symmetry \cite{He:2018uxa, Cordova:2018uop, Paulos:2018fym, Cordova:2019lot} this convex space was shown to have a vertex corresponding to the $O(N)$ integrable non-linear sigma model.

Similar considerations were applied to the case of resonances \cite{Doroud:2018szp} and multiple amplitudes \cite{Homrich:2019cbt}. Although in the case of $O(N)$ symmetry an integrable model was found at a vertex of the space, in other cases where there are integrable models with free parameters, a whole locus of the boundary of the space can correspond to an integrable model such is the case of SUSY, and $Z_2$ and $Z_4$ symmetries \cite{Bercini:2019vme} as well as theories with a boundary where one looks for the space of reflection matrices \cite{Kruczenski:2020ujw}.

Two spacetime dimensions is also where the \emph{dual} approach was pioneered in \cite{Cordova:2019lot}. The \emph{primal} approach of the aforementioned works always involves a truncation (and subsequent numerical exploration) of the space of scattering amplitudes. One therefore obtains extremal values, for example a largest possible coupling constant, that are always a bit smaller than the true extrema. On the other hand, in the dual approach one explores a space of constraints, leading to rigorous bounds and an approach of the true extrema from the other side. This latter approach is of course also widely in use in the numerical conformal bootstrap.

% The numerical work required to map out the space of S-matrices can be divided into two approaches, the primal and the dual approach. As we outline in more detail below, in the primal approach one solves analyticity and crossing by introducing a basis of functions that satisfy all these constraints. One then takes a finite-dimensional subset of these functions and imposes the (non-linear) unitarity constraints. In  one gradually approaches the boundary of the space from the inside. In the dual approach, one defines a space of Lagrange multipliers that impose the constraints. By gradually increasing the space of Lagrange multipliers more and more constraints are imposed and in that way one approaches the allowed space from outside. 

One important point of either approach is that the numerical algorithm not only estimates the extremal value for an observable but also the corresponding extremal amplitude. These amplitudes have been shown to agree with previously known analytic computations based on integrability or with other methods like Hamiltonian truncation \cite{Gabai:2019ryw}. It however appears that the amplitudes found at the boundary almost universally saturate the unitarity conditions (for all energies, up to numerical errors).\footnote{In the bootstrap study of multiple amplitudes \cite{Homrich:2019cbt}, it was found that unitarity was saturated among the two particle external states considered. Therefore, the extremal amplitudes are not always elastic.}
This implies that there is no particle creation, which is expected only in integrable theories. Therefore not all the extremal amplitudes  correspond to physical theories.

On the whole the S-matrix bootstrap approach can be considered a successful and interesting approach to 2d massive theories but the real test would be to obtain results in higher dimensions to which we now turn. 

\subsection{Higher-dimensional amplitudes }
\label{sec:D>2}
The modern non-perturbative S-matrix bootstrap in spacetime dimension $d>2$ started with the primal numerical algorithm proposed in \cite{Paulos:2017fhb}.
%Consider the 2 to 2 scattering amplitude of identical scalars on mass $m=1$.
%In this case, 
The algorithm has 4 main steps:\footnote{Here, we consider the simplest case of identical scalars of mass $m=1$ and no bound states. This applies to $\pi^0 - \pi^0$ elastic scattering.}
\begin{enumerate}
\item Write an ansatz that obeys analyticity, Lorentz invariance and crossing symmetry:
\beq
\label{Tansataz}
T={\rm poles} + \sum_{a+b+c\le N} \alpha_{(abc)} \rho_s^a \rho_t^b \rho_u^c
\eeq
where $\rho_s = \frac{q-\sqrt{4-s}}{q+\sqrt{4-s}}$ is a conformal mapping from the cut $s$-plane to the unit disk ($q>0$ is a constant).
\item Impose unitarity for each partial wave
\beq
\left| S_\ell(s) \right|^2 \le 1\,, \qquad s>4 \,,\qquad 0\le \ell\le L\,. 
\label{unitarity}
\eeq
\item Maximize a linear functional of the amplitude $T$ subject to the unitarity constraints. Notice that $S_\ell(s)$ is linear in $T$ and therefore in the parameters $\alpha_{(abc)}$. This makes the optimization problem semi-definite and we can use SDPB \cite{Simmons-Duffin:2015qma,Landry:2019qug}.
\item Extrapolate $L\to \infty$ (all constraints) and $N\to \infty$ (full space of analytic amplitudes). 
\end{enumerate}

In practice, this algorithm works and, after the extrapolation step, one effectively obtains bounds on a whole zoo of interesting observables like quartic and cubic couplings, scattering lengths, etc. The method can also be easily generalized to include global symmetries like isospin \cite{Guerrieri:2018uew,Bose:2020shm, Bose:2020cod}, and likewise extends to spinning particles as was done in 4 dimensions in \cite{Hebbar:2020ukp}. It is also possible to study or impose the presence of resonances. For such variations it is often convenient to modify the ansatz \eqref{Tansataz} consistent with analyticity and crossing. Sometimes these modifications are straightforward, e.g.\ when imposing a zero for a resonance, but in other cases they are more difficult. It is for example not yet known how to modify equation \eqref{Tansataz} to obtain Froissart-type behavior at high energies.

These intial successes warrant further exploration. One of the main open questions is the nature of the optimal amplitude in the limit $L\to \infty$ and $N\to \infty$. The numerical results suggest that as $N\to \infty$ the inequalities \eqref{unitarity} tend to be saturated for any $s$ and $\ell$. However, a purely elastic non-vanishing amplitude is not allowed \cite{Aks:1965qga,Correia:2020xtr}. It would be very interesting to avoid the numerical saturation of the unitarity constraints, and obtain physical amplitudes with non-zero particle creation. See \cite{Tourkine:2021fqh} for a first step in this direction.

%Double discontinuity and the necessity of inelasticity \cite{Correia:2020xtr}

%Relative entropy in scattering \cite{Bose:2020shm, Bose:2020cod}

%\subsection{ $D>2$ dual - Martin}
%\label{sec:dual}

The above constitutes the primal approach, which computes numerically a region contained inside the space of allowed S-matrices. There is also a higher-dimensional version of the aforementioned dual approach, which defines a region that contains the allowed space. This latter approach dates back to the work of Lopez and Mennesier \cite{Lopez:1976zs} in the 1970s and has been revived in \cite{He:2021eqn, Guerrieri:2021tak}. The work \cite{Guerrieri:2021tak} in particular exemplifies that Landau analyticity is not a necessity to set up a meaningful numerical program to constrain scattering amplitudes. Conceptually the dual approach has the nice feature that it is directly formulated in terms of (dual) partial waves instead of the amplitude. It is however far from given that either of the current approaches is optimal, both from a numerical and conceptual standpoint. In fact, in semidefinite programming problems there is always a dual version of each primal problem, where the role of Lagrange multipliers and fundamental variables is interchanged, and many solvers take a combined primal-dual approach.

% This novel approach provides immediately leads to a wide variety of ideas to explore. We discuss several specific directions below, but perhaps the main open question is how to obtain or parametrize physical amplitudes with non-zero particle creation. Although the ground work appears to be done, it seems that the most exciting results and specifically connections with experiment are still to come. 

% The combined primal-dual approach is standard in convex optimization problems and gives a nice conceptual framework for the S-matrix program.

% However, without using that language, these ideas were also part of the original S-matrix approach starting with the pioneer work of Lopez and Mennessier \cite{Lopez:1976zs}. Besides giving an alternative way to map the space of S-matrices, the dual approach has the advantage that it focuses on the Lagrange multipliers that impose the unitarity constraints and therefore is defined in terms of dual partial waves that are directly related to the physical partial waves accessible to experiment.

% Although the dual to the S-matrix bootstrap can be constructed \cite{He:2021eqn, Guerrieri:2021tak}, there are many ideas to explore, perhaps the main one is how to obtain S-matrices that have particle creation. Although the ground work appears to be done, it seems that the most exciting results and specially connections with experiment are still to come. 

\subsection{Massless particles}
\label{sec:EFT}
The numerical S-matrix Bootstrap can also be applied to massless particles, as long as their scattering amplitudes are non-perturbatively well defined (IR finite).
This can be used to place bounds on Wilson coefficients of the corresponding Effective Field Theories (EFTs).
The first application of this type was to flux tubes in confining gauge theories. More precisely, any two dimensional defect of a higher dimensional massive QFT that spontaneously breaks Poincare invariance is described by an EFT for the massless phonons (or branons) of the defect \cite{Aharony:2013ipa}. 
Following \cite{Dubovsky:2012sh}, the   two-dimensional S-matrix bootstrap approach has been applied to the two-to-two scattering amplitude of these branons, leading to non-perturbative bounds on the leading Wilson coefficients. In this case, there are analytic and numerical primal  
 \cite{EliasMiro:2019kyf} and dual \cite{EliasMiro:2021nul} results.
  One remarkable result of these works % is the presence of the world-sheet axion (as a resonance) with a coupling compatible with integrability \cite{}.
is that S-matrices living at the boundary of the allowed
space exhibit an intricate pattern of resonances with one sharper resonance whose quantum numbers, mass,
and width match those of the world-sheet axion previously observed in lattice simulations \cite{Athenodorou:2010cs, Dubovsky:2013gi, Dubovsky:2014fma, Athenodorou:2017cmw}.

%\JP{Add this? The lattice measurements of these Wilson coefficients are compatible with the bounds \cite{...}.}
 
In higher dimensions, there have been two salient applications of the numerical primal massless S-matrix bootstrap: pions \cite{Guerrieri:2020bto}
and supergravitons \cite{Guerrieri:2021ivu}.
  In this context, the first step of the algorithm described in \ref{sec:D>2} is modified to be compatible with the presence of massless particles and the low energy behaviour predicted by the EFT of interest.
In \cite{Guerrieri:2020bto}, the naive bounds of \cite{Pham:1985cr,Adams:2006sv} on the leading low energy constants of the chiral lagrangian were corrected including pion loop effects and the non-linear constraints from unitarity.
The paper \cite{Guerrieri:2021ivu} studied two-to-two scattering of supergravitons in 10 spacetime dimensions with maximal supersymmetry. Unitarity requires a non-trivial UV completion of supergravity and leads to a lower bound on the leading curvature correction in the effective action. Remarkably, this lower bound is very close to the lowest value realized in string theory when the string coupling is varied. This intriguing result may be interpreted as evidence for the \emph{String Lamppost Principle} \cite{Kim:2019ths}.

\subsection{UV information }
\label{sec:localop}

It would be very useful to input information about the UV behaviour of the QFT under study into the S-matrix bootstrap approach. This may aid to uniquely specify the theory. For example, it is possible to set up a mixed bootstrap which combines states created by acting with local operators on the vacuum with the usual scattering states \cite{Karateev:2019ymz, Karateev:2020axc}. The new terms in these mixed bootstrap equations are two-particle form factors and spectral densities of local operators, which both depend on a single variable in a well-understood analytic manner and therefore can be directly incorporated into the numerical primal approach.
In two spacetime dimensions,  this approach is able to probe the central charge of the UV CFT  \cite{Karateev:2019ymz}. For example, it was shown that the existence of $N$ massive particles transforming in the vector representation of an $O(N)$ global symmetry, implies a UV central charge scaling linearly with $N$.
In addition, it is possible to input spectral density and form factor data from numerical simulations of the Renormalization Group flow using, for example, Hamiltonian truncation methods. This approach was used to study the $\phi^4$ theory in 2d in \cite{Chen:2021pgx,Chen:2021bmm}.
For work similar in spirit see also \cite{Gabai:2019ryw} where the authors studied the 2d Ising field theory.

It would also be interesting to input hard scattering information \cite{Polchinski:2001tt} into the S-matrix Bootstrap. However, at the moment it is unclear how to make use of purely high energy properties like this.

\subsection{Unitarity versus Positivity}

The unitarity equations \eqref{unitarity} are often written in terms of the connected amplitude $f_\ell(s)=-i(S_\ell(s)-1)$, as follows
\beq
2\, {\rm Im}\, f_\ell(s) \ge | f_\ell(s)|^2\,.
\eeq
This implies the weaker condition $ {\rm Im}\, f_\ell(s) \ge 0$ known as \emph{positivity}. The main advantage of positivity with respect to unitarity is that it is a linear constraint.
Under the assumption of weak coupling unitarity reduces to positivity.

There has been a lot of recent progress in bounding EFT coefficients using positivity, see for example \cite{Bellazzini:2020cot, Tolley:2020gtv, Caron-Huot:2020cmc,  Sinha:2020win, Arkani-Hamed:2020blm, Li:2021lpe, Caron-Huot:2021rmr, Haldar:2021rri, Bern:2021ppb, Chiang:2021ziz, Henriksson:2021ymi, Chowdhury:2021ynh, Bellazzini:2021oaj, Caron-Huot:2022ugt}. Most of these  works assume a mass gap. This is justified under the assumption of weak coupling because one can neglect loop diagrams with massless particles. However, generically this is not the case and the usual methods of positivity break down. For example, the lower bound on curvature corrections to supergravity derived in \cite{Guerrieri:2021ivu} cannot be obtained solely with positivity.    Another interesting example is the leading non-universal correction $\gamma_3$ to the flux tube EFT in $D=3$. In this case, naive positivity   predicts $\gamma_3>0$ and the non-perturbative S-matrix bootstrap predicts $\gamma_3 > -\frac{1}{768}$ \cite{EliasMiro:2021nul}.
Only the latter is compatible with the lattice measurements for $SU(2)$ pure Yang-Mills \cite{Caristo:2021tbk}.

Confining gauge theories at the planar level  are an important open problem where unitarity reduces to positivity. In this case, the amplitudes are meromorphic functions with positive residues. In \cite{Caron-Huot:2016icg, Sever:2017ylk} it was shown that the (unphysical) regime
of $s,t \gg 1$ is universal. On the other hand, so far no useful constraints have been derived on the  mass spectrum and spins of the light particles.

It is also interesting to consider classical S-matrices for gravitons. 
In this case, an important open question is the uniqueness of Einstein gravity and the tree-level superstring amplitudes
\cite{Chowdhury:2019kaq, Huang:2020nqy},  
assuming the Classical Regge Growth conjecture
\cite{Chandorkar:2021viw}.

\subsection{QFT in AdS}
\label{sec:AdS}
An entirely orthogonal line of attack is based on studying quantum field theories in a fixed Anti-de Sitter (AdS) background. The focus is then on the boundary correlation functions, which are well-known from the AdS/CFT correspondence. These obey essentially all the axioms of CFT correlation functions, including for example a large and well-established domain of analyticity. Upon taking the flat-space limit, i.e. upon sending the AdS curvature radius $R$ to infinity, the boundary correlation functions should transmogrify into scattering amplitudes. For a gapped theory the flat-space limit implies that the scaling dimensions of boundary operators all become large. For example, for single-particle states we expect $ \Delta \sim m R$ with $m$ the mass of the bulk particle created by a boundary operator of dimension $\Delta$. The flat-space limit of the correlation functions can be taken in various ways. Suppressing normalization factors, we have the following prescriptions.
\begin{itemize}
   \item Using the Mellin amplitude $M(\gamma_{ij})$ for connected correlation functions introduced in \cite{Mack:2009mi}. For massless particles a prescription was introduced in \cite{Penedones:2010ue}, and for massive particles in \cite{Paulos:2016fap}; the latter simply reads:
  \begin{equation}
    T(k_i \cdot k_j) \propto \lim_{R \to \infty} M\left(\gamma_{ij} = m R ( 1 + k_i \cdot k_j/m^2)\right)
  \end{equation}
  so the Mellin amplitude directly becomes the scattering amplitude.
  \item In position space \cite{Dubovsky:2017cnj, Hijano:2019qmi, Hijano:2020szl, Komatsu:2020sag}. Again for massive particles, this prescription reads:
  \begin{equation}
  \vev{\underline{\tilde k}_1\ldots \underline{\tilde k}_a|S| \underline{k}_1 \ldots \underline{k}_b} \propto
\lim_{R \to \infty} 
\left. \langle\mathcal O(\tilde n_1) \ldots \mathcal O(\tilde n_a) \mathcal O(n_1) \ldots \mathcal O(n_b)\rangle \right|_{\text{S-matrix}}  
  \end{equation}
  where the precise map between boundary positions $n_i$ and in- and out-going on-shell momenta $k_i$ can be found in the above references.
  \item In spectral space. For example, the density $c(\Delta,\ell)$ obtained from the conformal partial wave decomposition reduces to the partial waves: 
  \begin{equation}
    f_\ell(s) \propto \lim_{R \to \infty} c_{\text{connected}}(\Delta = R \sqrt{s},\ell)\,.
  \end{equation}
  This is a variant of the phase shift formula of \cite{Paulos:2016fap}.
\end{itemize}
Certainly all these prescriptions are related, often by a saddle point approximation of the integral that transforms one right-hand side into the other. They however each have different regimes of validity and corresponding (dis)advantages.

One intriguing prospect is to use the above equations to derive rigorous results about scattering amplitudes using solely CFT axioms rather than, say, the axioms implied by the LSZ prescriptions. Indeed, suppose there exists a one-parameter family of solutions to the crossing equations labelled by $R$, then what can we say about the limit $R \to \infty$? Is the resulting object always an S-matrix? Would it always be unitary, analytic and crossing symmetric? Work in these directions is currently in progress by several groups.
 
One can also apply the familiar conformal bootstrap machinery directly to the boundary correlation functions, and extrapolate the resulting bounds to the flat-space limit. This was first done in \cite{Paulos:2016fap} where, for two-dimensional theories, an excellent match was found with the S-matrix bounds of \cite{Paulos:2016but}. The obvious advantage of this procedure, which is that it does not rely on unproven analyticity properties, is somewhat offset by the significant computational resources demanded by the QFT in AdS computations. The reason for this slowdown are the large operator dimensions appearing (for a gapped theory) when AdS is nearly flat, and in this regime the numerical bootstrap is less powerful. The problem was illustrated for three-dimensional theories in \cite{Paulos:2017fhb}. In this context it is therefore interesting to consider a different basis of functionals for the conformal bootstrap, which may speed up the numerical algorithms or even lead to analytic results. In two dimensions it was already shown in \cite{Mazac:2016qev, Mazac:2018mdx, Mazac:2018ycv} that alternative functionals can lead to a much more direct connection between the conformal and the S-matrix bootstrap.

The above results also marked the beginning of a revival of studies of QFT in AdS, with results ranging from perturbative computations (large $N$ models in \cite{Carmi:2018qzm}, weakly coupled RG flows in \cite{Giombi:2020rmc}) to general constraints (the $a$-theorem in AdS \cite{Kundu:2019zsl}, AdS Cutkosky rules in \cite{Meltzer:2020qbr}) to a study of numerical constraints for an entire RG flow in AdS \cite{Antunes:2021abs}. In the flat-space limit \cite{Haldar:2019prg} made a connection to the Froissart bound and \cite{Kundu:2020bdn} obtained connections on effective field theory coefficients in higher dimensions.

For a bulk theory with massless particles, as would be the case for gravitational theories, the works \cite{Caron-Huot:2021enk, Caron-Huot:2020adz} represent the state of the art. Here the authors demonstrate locality of the effective field theory in AdS dual to any CFT with a parametrically large central charge and spectral gap, and demonstrate analyticity of the amplitudes arising in the flat-space limit. These works provide an AdS version of the flat-space effective field theory bounds discussed above.

% CFT in AdS and boundary RG flows \cite{Giombi:2020rmc}
% Large N vector models 

% bulk a-anomaly from boundary CT \cite{Kundu:2019zsl, Kundu:2020bdn}

% Dalimil and Miguel did some analytic CFT bounds applied to QFT in AdS2 \cite{Mazac:2016qev, Mazac:2018mdx, Mazac:2018ycv}.

% CFT unitarity and the AdS Cutkosky rules \cite{Meltzer:2020qbr}

% Froissart bound: \cite{Haldar:2019prg}

%\subsection{Tree-level}
%\label{sec:tree}

% \subsection{Miscellanea - Balt}
% There are several research directions that we have not yet touched. One is the celestial sphere program which was reviewed in \cite{Pasterski:2021raf}. 

% This review may be useful \cite{Elvang:2020lue}.

% Probably we should mention the celestial sphere program \cite{Arkani-Hamed:2020gyp} 
% https://arxiv.org/pdf/2111.11392.pdf

% Tourkine + Zhiboedov fixed point iterative method.\BvR{Can we not put this in primal?}

\section{Future directions} 
Our vision is that the S-matrix bootstrap program will continue to develop along three different directions. First, there are surely many fundamental and universal properties of amplitudes and related bounds that remain to be uncovered. Second, there is the challenge of approximating the S-matrix for specific theories, i.e.~the non-perturbative completion of perturbative results that specify the UV behavior. Third, there is the opposite question of which IR EFTs admit a non-perturbative completion, which is of particular importance for gravitational theories. We elaborate and exemplify each of these directions below.

% See slides from Strings discussion

% 2 or 3 main future directions: 

\subsection{What is possible in QFT? }
In its modern incarnation, the S-matrix bootstrap aims to constrain (rather than directly solve) physical observables. This naturally leads to a research program that seeks to obtain significant ``axiomatic constraints'' on scattering amplitudes, irrespective of whether physical theories saturate these constraints. So far we know of a large variety of constraints like bounds on scattering lengths, couplings, the value of the scattering amplitude in the Euclidean domain, and so on. But what else is possible? Can we say more about large spin partial waves, as is done for the Froissart bound or for Aks' theorem? Can we concretize the predictions of Regge theory? 
It would be fascinating to see what the methods sketched above can teach us.

\subsection{Strongly coupled gauge theories  }
One of the main problems in physics is to understand the confining regime of non-abelian gauge theories, in particular QCD. In fact understanding the strong interaction was the objective of the original S-matrix bootstrap. Following the new ideas one would like to map out the space of possible low energy theories and perhaps find QCD at a distinguished point in that space as happened in the 2d case. Although progress has been made in that direction by studying pion scattering (see for example \cite{Guerrieri:2018uew,Guerrieri:2020bto}), there is a lot to be explored in this area. 
For example, it seems feasible to  bootstrap multiple two-to-two amplitudes involving pions and nucleons (assuming exact isospin symmetry).
  Can we study higher particle amplitudes? This would also allow to impose the soft relations that follow from the chiral lagrangian.
Perhaps the main problem is how to incorporate UV information that distinguishes QCD from other  QFTs with similar low energy spectrum. 

In 2+1 dimensions, the dichotomy between bosons and fermions breaks down and particles can be anyons \cite{Wilczek:1982wy}. It would be very interesting to study their scattering amplitudes.
Explicit computations in Chern-Simons-matter theories \cite{Jain:2014nza} show that the crossing equations
need to be modified to accommodate the non-trivial phases of anyon statistics. Nevertheless,
a systematic relativistic S-matrix theory of anyons is still lacking.

\subsection{Quantum gravity  }
In the context of gravitational scattering there are many open questions for the future:
\begin{itemize}
\item Bound the leading higher curvature corrections to maximal supergravity in different spacetime dimensions ($5 \le D \le 11$). In particular, there is a unique prediction from M-theory in $D = 11$ which  would be interesting to compare with the numerical bootstrap results.
\item Study the allowed space of Wilson coefficients including subleading corrections in $D = 10$ and $D = 11$ where there are concrete predictions from Superstring Theory.
\item  Study the curvature corrections without supersymmetry. This will require setting up the S-matrix Bootstrap for spinning particles in different spacetime dimensions, generalizing \cite{Hebbar:2020ukp}.
\item Study the effect of inelasticity (e.g. from black hole production) on the S-matrix Bootstrap bounds.
\item Bootstrap IR safe observables in 4D. This involves defining finite   observables (e.g. following the classic paper \cite{Kulish:1970ut} or using celestial amplitudes \cite{Arkani-Hamed:2020gyp}) and understanding their analyticity and unitarity properties. Similar comments apply to scattering of charged particles and photons in 4D.
\end{itemize}

In practice, the most urgent improvement is the development of a dual S-matrix Bootstrap approach to scattering of massless particles in more than 2 spacetime dimensions.
A more efficient numerical algorithm will allow us to more easily explore many of the open problems listed above.

\section{Conclusion   }
\label{sec:conclusion}
The S-matrix bootstrap from the 1950s and 1960s, which aimed at defining the S-matrix from its analyticity, crossing, and symmetry properties, has been reborn as a tool to understand quantum field theories by carving out the space of possible S-matrices. In certain cases we can additionally use properties of the theory to identify some distinguished point corresponding to the theory in question. Combined with new numerical optimization methods and computational capabilities we obtain a new powerful numerical technique to explore scattering amplitudes for strongly coupled theories. We think the most exciting results of this investigation are yet to come, and eagerly await the new insights it will produce into both strongly coupled gauge theories and gravity.  

\section*{Acknowledgments}
We would like to thank Andrea Guerrieri, Yifei He, Denis Karateev, Miguel Paulos, Aninda Sinha, Pedro Vieira, Sasha Zhiboedov for discussion and comments on the draft. We are supported by Simons Foundation grants \#488649 (JP) and \#488659 (BvR) (Simons collaboration on the non-perturbative bootstrap). JP is supported by the Swiss National Science Foundation through the project 200020\_197160 and through the National Centre of Competence in Research SwissMAP. In addition, MK is very grateful to the DOE that supported in part this work through grant DE-SC0007884 and the QuantiSED Fermilab consortium, as well as to the Keck Foundation that also provided partial support for this work.

\bibliography{whitepaper} 

\providecommand{\noopsort}[1]{}\providecommand{\singleletter}[1]{#1}%
\begin{thebibliography}{93}
\expandafter\ifx\csname natexlab\endcsname\relax\def\natexlab#1{#1}\fi
\expandafter\ifx\csname bibnamefont\endcsname\relax
  \def\bibnamefont#1{#1}\fi
\expandafter\ifx\csname bibfnamefont\endcsname\relax
  \def\bibfnamefont#1{#1}\fi
\expandafter\ifx\csname citenamefont\endcsname\relax
  \def\citenamefont#1{#1}\fi
\expandafter\ifx\csname url\endcsname\relax
  \def\url#1{\texttt{#1}}\fi
\expandafter\ifx\csname urlprefix\endcsname\relax\def\urlprefix{URL }\fi
\providecommand{\bibinfo}[2]{#2}
\providecommand{\eprint}[2][]{\url{#2}}

\bibitem[{\citenamefont{Bros et~al.}(1965)\citenamefont{Bros, Epstein, and
  Glaser}}]{Bros:1965kbd}
\bibinfo{author}{\bibfnamefont{J.}~\bibnamefont{Bros}},
  \bibinfo{author}{\bibfnamefont{H.}~\bibnamefont{Epstein}}, \bibnamefont{and}
  \bibinfo{author}{\bibfnamefont{V.}~\bibnamefont{Glaser}},
  \bibinfo{journal}{Commun. Math. Phys.} \textbf{\bibinfo{volume}{1}},
  \bibinfo{pages}{240} (\bibinfo{year}{1965}).

\bibitem[{\citenamefont{Martin}(1965)}]{Martin:1965jj}
\bibinfo{author}{\bibfnamefont{A.}~\bibnamefont{Martin}},
  \bibinfo{journal}{Nuovo Cim. A} \textbf{\bibinfo{volume}{42}},
  \bibinfo{pages}{930} (\bibinfo{year}{1965}).

\bibitem[{\citenamefont{Sommer}(1970)}]{Sommer:1970mr}
\bibinfo{author}{\bibfnamefont{G.}~\bibnamefont{Sommer}},
  \bibinfo{journal}{Fortsch. Phys.} \textbf{\bibinfo{volume}{18}},
  \bibinfo{pages}{577} (\bibinfo{year}{1970}).

\bibitem[{\citenamefont{Bogolyubov et~al.}(1990)\citenamefont{Bogolyubov,
  Logunov, Oksak, and Todorov}}]{Bogolyubov:1990kw}
\bibinfo{author}{\bibfnamefont{N.~N.} \bibnamefont{Bogolyubov}},
  \bibinfo{author}{\bibfnamefont{A.~A.} \bibnamefont{Logunov}},
  \bibinfo{author}{\bibfnamefont{A.~I.} \bibnamefont{Oksak}}, \bibnamefont{and}
  \bibinfo{author}{\bibfnamefont{I.~T.} \bibnamefont{Todorov}},
  \emph{\bibinfo{title}{{General principles of quantum field theory}}}
  (\bibinfo{year}{1990}).

\bibitem[{\citenamefont{Correia et~al.}(2020)\citenamefont{Correia, Sever, and
  Zhiboedov}}]{Correia:2020xtr}
\bibinfo{author}{\bibfnamefont{M.}~\bibnamefont{Correia}},
  \bibinfo{author}{\bibfnamefont{A.}~\bibnamefont{Sever}}, \bibnamefont{and}
  \bibinfo{author}{\bibfnamefont{A.}~\bibnamefont{Zhiboedov}}
  (\bibinfo{year}{2020}), \eprint{2006.08221}.

\bibitem[{\citenamefont{Coleman and Thun}(1978)}]{Coleman:1978kk}
\bibinfo{author}{\bibfnamefont{S.~R.} \bibnamefont{Coleman}} \bibnamefont{and}
  \bibinfo{author}{\bibfnamefont{H.~J.} \bibnamefont{Thun}},
  \bibinfo{journal}{Commun. Math. Phys.} \textbf{\bibinfo{volume}{61}},
  \bibinfo{pages}{31} (\bibinfo{year}{1978}).

\bibitem[{\citenamefont{Mizera}(2021{\natexlab{a}})}]{Mizera:2021ujs}
\bibinfo{author}{\bibfnamefont{S.}~\bibnamefont{Mizera}},
  \bibinfo{journal}{Phys. Rev. D} \textbf{\bibinfo{volume}{103}},
  \bibinfo{pages}{081701} (\bibinfo{year}{2021}{\natexlab{a}}),
  \eprint{2101.08266}.

\bibitem[{\citenamefont{Mizera}(2021{\natexlab{b}})}]{Mizera:2021fap}
\bibinfo{author}{\bibfnamefont{S.}~\bibnamefont{Mizera}},
  \bibinfo{journal}{Phys. Rev. D} \textbf{\bibinfo{volume}{104}},
  \bibinfo{pages}{045003} (\bibinfo{year}{2021}{\natexlab{b}}),
  \eprint{2104.12776}.

\bibitem[{\citenamefont{Correia et~al.}(2021)\citenamefont{Correia, Sever, and
  Zhiboedov}}]{Correia:2021etg}
\bibinfo{author}{\bibfnamefont{M.}~\bibnamefont{Correia}},
  \bibinfo{author}{\bibfnamefont{A.}~\bibnamefont{Sever}}, \bibnamefont{and}
  \bibinfo{author}{\bibfnamefont{A.}~\bibnamefont{Zhiboedov}}
  (\bibinfo{year}{2021}), \eprint{2111.12100}.

\bibitem[{\citenamefont{Paulos et~al.}(2017{\natexlab{a}})\citenamefont{Paulos,
  Penedones, Toledo, van Rees, and Vieira}}]{Paulos:2016but}
\bibinfo{author}{\bibfnamefont{M.~F.} \bibnamefont{Paulos}},
  \bibinfo{author}{\bibfnamefont{J.}~\bibnamefont{Penedones}},
  \bibinfo{author}{\bibfnamefont{J.}~\bibnamefont{Toledo}},
  \bibinfo{author}{\bibfnamefont{B.~C.} \bibnamefont{van Rees}},
  \bibnamefont{and} \bibinfo{author}{\bibfnamefont{P.}~\bibnamefont{Vieira}},
  \bibinfo{journal}{JHEP} \textbf{\bibinfo{volume}{11}}, \bibinfo{pages}{143}
  (\bibinfo{year}{2017}{\natexlab{a}}), \eprint{1607.06110}.

\bibitem[{\citenamefont{Creutz}(1972)}]{Creutz:1973rw}
\bibinfo{author}{\bibfnamefont{M.}~\bibnamefont{Creutz}},
  \bibinfo{journal}{Phys. Rev. D} \textbf{\bibinfo{volume}{6}},
  \bibinfo{pages}{2763} (\bibinfo{year}{1972}).

\bibitem[{\citenamefont{He et~al.}(2018)\citenamefont{He, Irrgang, and
  Kruczenski}}]{He:2018uxa}
\bibinfo{author}{\bibfnamefont{Y.}~\bibnamefont{He}},
  \bibinfo{author}{\bibfnamefont{A.}~\bibnamefont{Irrgang}}, \bibnamefont{and}
  \bibinfo{author}{\bibfnamefont{M.}~\bibnamefont{Kruczenski}},
  \bibinfo{journal}{JHEP} \textbf{\bibinfo{volume}{11}}, \bibinfo{pages}{093}
  (\bibinfo{year}{2018}), \eprint{1805.02812}.

\bibitem[{\citenamefont{C\'ordova and Vieira}(2018)}]{Cordova:2018uop}
\bibinfo{author}{\bibfnamefont{L.}~\bibnamefont{C\'ordova}} \bibnamefont{and}
  \bibinfo{author}{\bibfnamefont{P.}~\bibnamefont{Vieira}},
  \bibinfo{journal}{JHEP} \textbf{\bibinfo{volume}{12}}, \bibinfo{pages}{063}
  (\bibinfo{year}{2018}), \eprint{1805.11143}.

\bibitem[{\citenamefont{Paulos and Zheng}(2020)}]{Paulos:2018fym}
\bibinfo{author}{\bibfnamefont{M.~F.} \bibnamefont{Paulos}} \bibnamefont{and}
  \bibinfo{author}{\bibfnamefont{Z.}~\bibnamefont{Zheng}},
  \bibinfo{journal}{JHEP} \textbf{\bibinfo{volume}{05}}, \bibinfo{pages}{145}
  (\bibinfo{year}{2020}), \eprint{1805.11429}.

\bibitem[{\citenamefont{C\'ordova et~al.}(2020)\citenamefont{C\'ordova, He,
  Kruczenski, and Vieira}}]{Cordova:2019lot}
\bibinfo{author}{\bibfnamefont{L.}~\bibnamefont{C\'ordova}},
  \bibinfo{author}{\bibfnamefont{Y.}~\bibnamefont{He}},
  \bibinfo{author}{\bibfnamefont{M.}~\bibnamefont{Kruczenski}},
  \bibnamefont{and} \bibinfo{author}{\bibfnamefont{P.}~\bibnamefont{Vieira}},
  \bibinfo{journal}{JHEP} \textbf{\bibinfo{volume}{04}}, \bibinfo{pages}{142}
  (\bibinfo{year}{2020}), \eprint{1909.06495}.

\bibitem[{\citenamefont{Doroud and Elias~Mir\'o}(2018)}]{Doroud:2018szp}
\bibinfo{author}{\bibfnamefont{N.}~\bibnamefont{Doroud}} \bibnamefont{and}
  \bibinfo{author}{\bibfnamefont{J.}~\bibnamefont{Elias~Mir\'o}},
  \bibinfo{journal}{JHEP} \textbf{\bibinfo{volume}{09}}, \bibinfo{pages}{052}
  (\bibinfo{year}{2018}), \eprint{1804.04376}.

\bibitem[{\citenamefont{Homrich et~al.}(2019)\citenamefont{Homrich, Penedones,
  Toledo, van Rees, and Vieira}}]{Homrich:2019cbt}
\bibinfo{author}{\bibfnamefont{A.}~\bibnamefont{Homrich}},
  \bibinfo{author}{\bibfnamefont{J.}~\bibnamefont{Penedones}},
  \bibinfo{author}{\bibfnamefont{J.}~\bibnamefont{Toledo}},
  \bibinfo{author}{\bibfnamefont{B.~C.} \bibnamefont{van Rees}},
  \bibnamefont{and} \bibinfo{author}{\bibfnamefont{P.}~\bibnamefont{Vieira}},
  \bibinfo{journal}{JHEP} \textbf{\bibinfo{volume}{11}}, \bibinfo{pages}{076}
  (\bibinfo{year}{2019}), \eprint{1905.06905}.

\bibitem[{\citenamefont{Bercini et~al.}(2020)\citenamefont{Bercini, Fabri,
  Homrich, and Vieira}}]{Bercini:2019vme}
\bibinfo{author}{\bibfnamefont{C.}~\bibnamefont{Bercini}},
  \bibinfo{author}{\bibfnamefont{M.}~\bibnamefont{Fabri}},
  \bibinfo{author}{\bibfnamefont{A.}~\bibnamefont{Homrich}}, \bibnamefont{and}
  \bibinfo{author}{\bibfnamefont{P.}~\bibnamefont{Vieira}},
  \bibinfo{journal}{Phys. Rev. D} \textbf{\bibinfo{volume}{101}},
  \bibinfo{pages}{045022} (\bibinfo{year}{2020}), \eprint{1909.06453}.

\bibitem[{\citenamefont{Kruczenski and Murali}(2020)}]{Kruczenski:2020ujw}
\bibinfo{author}{\bibfnamefont{M.}~\bibnamefont{Kruczenski}} \bibnamefont{and}
  \bibinfo{author}{\bibfnamefont{H.}~\bibnamefont{Murali}}
  (\bibinfo{year}{2020}), \eprint{2012.15576}.

\bibitem[{\citenamefont{Gabai and Yin}(2019)}]{Gabai:2019ryw}
\bibinfo{author}{\bibfnamefont{B.}~\bibnamefont{Gabai}} \bibnamefont{and}
  \bibinfo{author}{\bibfnamefont{X.}~\bibnamefont{Yin}} (\bibinfo{year}{2019}),
  \eprint{1905.00710}.

\bibitem[{\citenamefont{Paulos et~al.}(2019)\citenamefont{Paulos, Penedones,
  Toledo, van Rees, and Vieira}}]{Paulos:2017fhb}
\bibinfo{author}{\bibfnamefont{M.~F.} \bibnamefont{Paulos}},
  \bibinfo{author}{\bibfnamefont{J.}~\bibnamefont{Penedones}},
  \bibinfo{author}{\bibfnamefont{J.}~\bibnamefont{Toledo}},
  \bibinfo{author}{\bibfnamefont{B.~C.} \bibnamefont{van Rees}},
  \bibnamefont{and} \bibinfo{author}{\bibfnamefont{P.}~\bibnamefont{Vieira}},
  \bibinfo{journal}{JHEP} \textbf{\bibinfo{volume}{12}}, \bibinfo{pages}{040}
  (\bibinfo{year}{2019}), \eprint{1708.06765}.

\bibitem[{\citenamefont{Simmons-Duffin}(2015)}]{Simmons-Duffin:2015qma}
\bibinfo{author}{\bibfnamefont{D.}~\bibnamefont{Simmons-Duffin}},
  \bibinfo{journal}{JHEP} \textbf{\bibinfo{volume}{06}}, \bibinfo{pages}{174}
  (\bibinfo{year}{2015}), \eprint{1502.02033}.

\bibitem[{\citenamefont{Landry and Simmons-Duffin}(2019)}]{Landry:2019qug}
\bibinfo{author}{\bibfnamefont{W.}~\bibnamefont{Landry}} \bibnamefont{and}
  \bibinfo{author}{\bibfnamefont{D.}~\bibnamefont{Simmons-Duffin}}
  (\bibinfo{year}{2019}), \eprint{1909.09745}.

\bibitem[{\citenamefont{Guerrieri et~al.}(2019)\citenamefont{Guerrieri,
  Penedones, and Vieira}}]{Guerrieri:2018uew}
\bibinfo{author}{\bibfnamefont{A.~L.} \bibnamefont{Guerrieri}},
  \bibinfo{author}{\bibfnamefont{J.}~\bibnamefont{Penedones}},
  \bibnamefont{and} \bibinfo{author}{\bibfnamefont{P.}~\bibnamefont{Vieira}},
  \bibinfo{journal}{Phys. Rev. Lett.} \textbf{\bibinfo{volume}{122}},
  \bibinfo{pages}{241604} (\bibinfo{year}{2019}), \eprint{1810.12849}.

\bibitem[{\citenamefont{Bose et~al.}(2020{\natexlab{a}})\citenamefont{Bose,
  Haldar, Sinha, Sinha, and Tiwari}}]{Bose:2020shm}
\bibinfo{author}{\bibfnamefont{A.}~\bibnamefont{Bose}},
  \bibinfo{author}{\bibfnamefont{P.}~\bibnamefont{Haldar}},
  \bibinfo{author}{\bibfnamefont{A.}~\bibnamefont{Sinha}},
  \bibinfo{author}{\bibfnamefont{P.}~\bibnamefont{Sinha}}, \bibnamefont{and}
  \bibinfo{author}{\bibfnamefont{S.~S.} \bibnamefont{Tiwari}},
  \bibinfo{journal}{SciPost Phys.} \textbf{\bibinfo{volume}{9}},
  \bibinfo{pages}{081} (\bibinfo{year}{2020}{\natexlab{a}}),
  \eprint{2006.12213}.

\bibitem[{\citenamefont{Bose et~al.}(2020{\natexlab{b}})\citenamefont{Bose,
  Sinha, and Tiwari}}]{Bose:2020cod}
\bibinfo{author}{\bibfnamefont{A.}~\bibnamefont{Bose}},
  \bibinfo{author}{\bibfnamefont{A.}~\bibnamefont{Sinha}}, \bibnamefont{and}
  \bibinfo{author}{\bibfnamefont{S.~S.} \bibnamefont{Tiwari}}
  (\bibinfo{year}{2020}{\natexlab{b}}), \eprint{2011.07944}.

\bibitem[{\citenamefont{Hebbar et~al.}(2020)\citenamefont{Hebbar, Karateev, and
  Penedones}}]{Hebbar:2020ukp}
\bibinfo{author}{\bibfnamefont{A.}~\bibnamefont{Hebbar}},
  \bibinfo{author}{\bibfnamefont{D.}~\bibnamefont{Karateev}}, \bibnamefont{and}
  \bibinfo{author}{\bibfnamefont{J.}~\bibnamefont{Penedones}}
  (\bibinfo{year}{2020}), \eprint{2011.11708}.

\bibitem[{\citenamefont{Aks}(1965)}]{Aks:1965qga}
\bibinfo{author}{\bibfnamefont{S.~O.} \bibnamefont{Aks}}, \bibinfo{journal}{J.
  Math. Phys.} \textbf{\bibinfo{volume}{6}}, \bibinfo{pages}{516}
  (\bibinfo{year}{1965}).

\bibitem[{\citenamefont{Tourkine and Zhiboedov}(2021)}]{Tourkine:2021fqh}
\bibinfo{author}{\bibfnamefont{P.}~\bibnamefont{Tourkine}} \bibnamefont{and}
  \bibinfo{author}{\bibfnamefont{A.}~\bibnamefont{Zhiboedov}},
  \bibinfo{journal}{JHEP} \textbf{\bibinfo{volume}{07}}, \bibinfo{pages}{228}
  (\bibinfo{year}{2021}), \eprint{2101.05211}.

\bibitem[{\citenamefont{Lopez and Mennessier}(1977)}]{Lopez:1976zs}
\bibinfo{author}{\bibfnamefont{C.}~\bibnamefont{Lopez}} \bibnamefont{and}
  \bibinfo{author}{\bibfnamefont{G.}~\bibnamefont{Mennessier}},
  \bibinfo{journal}{Nucl. Phys. B} \textbf{\bibinfo{volume}{118}},
  \bibinfo{pages}{426} (\bibinfo{year}{1977}).

\bibitem[{\citenamefont{He and Kruczenski}(2021)}]{He:2021eqn}
\bibinfo{author}{\bibfnamefont{Y.}~\bibnamefont{He}} \bibnamefont{and}
  \bibinfo{author}{\bibfnamefont{M.}~\bibnamefont{Kruczenski}},
  \bibinfo{journal}{JHEP} \textbf{\bibinfo{volume}{08}}, \bibinfo{pages}{125}
  (\bibinfo{year}{2021}), \eprint{2103.11484}.

\bibitem[{\citenamefont{Guerrieri and Sever}(2021)}]{Guerrieri:2021tak}
\bibinfo{author}{\bibfnamefont{A.}~\bibnamefont{Guerrieri}} \bibnamefont{and}
  \bibinfo{author}{\bibfnamefont{A.}~\bibnamefont{Sever}},
  \bibinfo{journal}{Phys. Rev. Lett.} \textbf{\bibinfo{volume}{127}},
  \bibinfo{pages}{251601} (\bibinfo{year}{2021}), \eprint{2106.10257}.

\bibitem[{\citenamefont{Aharony and Komargodski}(2013)}]{Aharony:2013ipa}
\bibinfo{author}{\bibfnamefont{O.}~\bibnamefont{Aharony}} \bibnamefont{and}
  \bibinfo{author}{\bibfnamefont{Z.}~\bibnamefont{Komargodski}},
  \bibinfo{journal}{JHEP} \textbf{\bibinfo{volume}{05}}, \bibinfo{pages}{118}
  (\bibinfo{year}{2013}), \eprint{1302.6257}.

\bibitem[{\citenamefont{Dubovsky et~al.}(2012)\citenamefont{Dubovsky, Flauger,
  and Gorbenko}}]{Dubovsky:2012sh}
\bibinfo{author}{\bibfnamefont{S.}~\bibnamefont{Dubovsky}},
  \bibinfo{author}{\bibfnamefont{R.}~\bibnamefont{Flauger}}, \bibnamefont{and}
  \bibinfo{author}{\bibfnamefont{V.}~\bibnamefont{Gorbenko}},
  \bibinfo{journal}{JHEP} \textbf{\bibinfo{volume}{09}}, \bibinfo{pages}{044}
  (\bibinfo{year}{2012}), \eprint{1203.1054}.

\bibitem[{\citenamefont{Elias~Mir\'o et~al.}(2019)\citenamefont{Elias~Mir\'o,
  Guerrieri, Hebbar, Penedones, and Vieira}}]{EliasMiro:2019kyf}
\bibinfo{author}{\bibfnamefont{J.}~\bibnamefont{Elias~Mir\'o}},
  \bibinfo{author}{\bibfnamefont{A.~L.} \bibnamefont{Guerrieri}},
  \bibinfo{author}{\bibfnamefont{A.}~\bibnamefont{Hebbar}},
  \bibinfo{author}{\bibfnamefont{J.}~\bibnamefont{Penedones}},
  \bibnamefont{and} \bibinfo{author}{\bibfnamefont{P.}~\bibnamefont{Vieira}},
  \bibinfo{journal}{Phys. Rev. Lett.} \textbf{\bibinfo{volume}{123}},
  \bibinfo{pages}{221602} (\bibinfo{year}{2019}), \eprint{1906.08098}.

\bibitem[{\citenamefont{Elias~Mir\'o and Guerrieri}(2021)}]{EliasMiro:2021nul}
\bibinfo{author}{\bibfnamefont{J.}~\bibnamefont{Elias~Mir\'o}}
  \bibnamefont{and}
  \bibinfo{author}{\bibfnamefont{A.}~\bibnamefont{Guerrieri}},
  \bibinfo{journal}{JHEP} \textbf{\bibinfo{volume}{10}}, \bibinfo{pages}{126}
  (\bibinfo{year}{2021}), \eprint{2106.07957}.

\bibitem[{\citenamefont{Athenodorou et~al.}(2011)\citenamefont{Athenodorou,
  Bringoltz, and Teper}}]{Athenodorou:2010cs}
\bibinfo{author}{\bibfnamefont{A.}~\bibnamefont{Athenodorou}},
  \bibinfo{author}{\bibfnamefont{B.}~\bibnamefont{Bringoltz}},
  \bibnamefont{and} \bibinfo{author}{\bibfnamefont{M.}~\bibnamefont{Teper}},
  \bibinfo{journal}{JHEP} \textbf{\bibinfo{volume}{02}}, \bibinfo{pages}{030}
  (\bibinfo{year}{2011}), \eprint{1007.4720}.

\bibitem[{\citenamefont{Dubovsky et~al.}(2013)\citenamefont{Dubovsky, Flauger,
  and Gorbenko}}]{Dubovsky:2013gi}
\bibinfo{author}{\bibfnamefont{S.}~\bibnamefont{Dubovsky}},
  \bibinfo{author}{\bibfnamefont{R.}~\bibnamefont{Flauger}}, \bibnamefont{and}
  \bibinfo{author}{\bibfnamefont{V.}~\bibnamefont{Gorbenko}},
  \bibinfo{journal}{Phys. Rev. Lett.} \textbf{\bibinfo{volume}{111}},
  \bibinfo{pages}{062006} (\bibinfo{year}{2013}), \eprint{1301.2325}.

\bibitem[{\citenamefont{Dubovsky et~al.}(2015)\citenamefont{Dubovsky, Flauger,
  and Gorbenko}}]{Dubovsky:2014fma}
\bibinfo{author}{\bibfnamefont{S.}~\bibnamefont{Dubovsky}},
  \bibinfo{author}{\bibfnamefont{R.}~\bibnamefont{Flauger}}, \bibnamefont{and}
  \bibinfo{author}{\bibfnamefont{V.}~\bibnamefont{Gorbenko}},
  \bibinfo{journal}{J. Exp. Theor. Phys.} \textbf{\bibinfo{volume}{120}},
  \bibinfo{pages}{399} (\bibinfo{year}{2015}), \eprint{1404.0037}.

\bibitem[{\citenamefont{Athenodorou and Teper}(2017)}]{Athenodorou:2017cmw}
\bibinfo{author}{\bibfnamefont{A.}~\bibnamefont{Athenodorou}} \bibnamefont{and}
  \bibinfo{author}{\bibfnamefont{M.}~\bibnamefont{Teper}},
  \bibinfo{journal}{Phys. Lett. B} \textbf{\bibinfo{volume}{771}},
  \bibinfo{pages}{408} (\bibinfo{year}{2017}), \eprint{1702.03717}.

\bibitem[{\citenamefont{Guerrieri et~al.}(2020)\citenamefont{Guerrieri,
  Penedones, and Vieira}}]{Guerrieri:2020bto}
\bibinfo{author}{\bibfnamefont{A.}~\bibnamefont{Guerrieri}},
  \bibinfo{author}{\bibfnamefont{J.}~\bibnamefont{Penedones}},
  \bibnamefont{and} \bibinfo{author}{\bibfnamefont{P.}~\bibnamefont{Vieira}}
  (\bibinfo{year}{2020}), \eprint{2011.02802}.

\bibitem[{\citenamefont{Guerrieri et~al.}(2021)\citenamefont{Guerrieri,
  Penedones, and Vieira}}]{Guerrieri:2021ivu}
\bibinfo{author}{\bibfnamefont{A.}~\bibnamefont{Guerrieri}},
  \bibinfo{author}{\bibfnamefont{J.}~\bibnamefont{Penedones}},
  \bibnamefont{and} \bibinfo{author}{\bibfnamefont{P.}~\bibnamefont{Vieira}},
  \bibinfo{journal}{Phys. Rev. Lett.} \textbf{\bibinfo{volume}{127}},
  \bibinfo{pages}{081601} (\bibinfo{year}{2021}), \eprint{2102.02847}.

\bibitem[{\citenamefont{Pham and Truong}(1985)}]{Pham:1985cr}
\bibinfo{author}{\bibfnamefont{T.~N.} \bibnamefont{Pham}} \bibnamefont{and}
  \bibinfo{author}{\bibfnamefont{T.~N.} \bibnamefont{Truong}},
  \bibinfo{journal}{Phys. Rev. D} \textbf{\bibinfo{volume}{31}},
  \bibinfo{pages}{3027} (\bibinfo{year}{1985}).

\bibitem[{\citenamefont{Adams et~al.}(2006)\citenamefont{Adams, Arkani-Hamed,
  Dubovsky, Nicolis, and Rattazzi}}]{Adams:2006sv}
\bibinfo{author}{\bibfnamefont{A.}~\bibnamefont{Adams}},
  \bibinfo{author}{\bibfnamefont{N.}~\bibnamefont{Arkani-Hamed}},
  \bibinfo{author}{\bibfnamefont{S.}~\bibnamefont{Dubovsky}},
  \bibinfo{author}{\bibfnamefont{A.}~\bibnamefont{Nicolis}}, \bibnamefont{and}
  \bibinfo{author}{\bibfnamefont{R.}~\bibnamefont{Rattazzi}},
  \bibinfo{journal}{JHEP} \textbf{\bibinfo{volume}{10}}, \bibinfo{pages}{014}
  (\bibinfo{year}{2006}), \eprint{hep-th/0602178}.

\bibitem[{\citenamefont{Kim et~al.}(2020)\citenamefont{Kim, Tarazi, and
  Vafa}}]{Kim:2019ths}
\bibinfo{author}{\bibfnamefont{H.-C.} \bibnamefont{Kim}},
  \bibinfo{author}{\bibfnamefont{H.-C.} \bibnamefont{Tarazi}},
  \bibnamefont{and} \bibinfo{author}{\bibfnamefont{C.}~\bibnamefont{Vafa}},
  \bibinfo{journal}{Phys. Rev. D} \textbf{\bibinfo{volume}{102}},
  \bibinfo{pages}{026003} (\bibinfo{year}{2020}), \eprint{1912.06144}.

\bibitem[{\citenamefont{Karateev et~al.}(2020)\citenamefont{Karateev, Kuhn, and
  Penedones}}]{Karateev:2019ymz}
\bibinfo{author}{\bibfnamefont{D.}~\bibnamefont{Karateev}},
  \bibinfo{author}{\bibfnamefont{S.}~\bibnamefont{Kuhn}}, \bibnamefont{and}
  \bibinfo{author}{\bibfnamefont{J.}~\bibnamefont{Penedones}},
  \bibinfo{journal}{JHEP} \textbf{\bibinfo{volume}{07}}, \bibinfo{pages}{035}
  (\bibinfo{year}{2020}), \eprint{1912.08940}.

\bibitem[{\citenamefont{Karateev}(2020)}]{Karateev:2020axc}
\bibinfo{author}{\bibfnamefont{D.}~\bibnamefont{Karateev}}
  (\bibinfo{year}{2020}), \eprint{2012.08538}.

\bibitem[{\citenamefont{Chen et~al.}(2022{\natexlab{a}})\citenamefont{Chen,
  Fitzpatrick, and Karateev}}]{Chen:2021pgx}
\bibinfo{author}{\bibfnamefont{H.}~\bibnamefont{Chen}},
  \bibinfo{author}{\bibfnamefont{A.~L.} \bibnamefont{Fitzpatrick}},
  \bibnamefont{and} \bibinfo{author}{\bibfnamefont{D.}~\bibnamefont{Karateev}},
  \bibinfo{journal}{JHEP} \textbf{\bibinfo{volume}{02}}, \bibinfo{pages}{146}
  (\bibinfo{year}{2022}{\natexlab{a}}), \eprint{2107.10286}.

\bibitem[{\citenamefont{Chen et~al.}(2022{\natexlab{b}})\citenamefont{Chen,
  Fitzpatrick, and Karateev}}]{Chen:2021bmm}
\bibinfo{author}{\bibfnamefont{H.}~\bibnamefont{Chen}},
  \bibinfo{author}{\bibfnamefont{A.~L.} \bibnamefont{Fitzpatrick}},
  \bibnamefont{and} \bibinfo{author}{\bibfnamefont{D.}~\bibnamefont{Karateev}},
  \bibinfo{journal}{JHEP} \textbf{\bibinfo{volume}{04}}, \bibinfo{pages}{109}
  (\bibinfo{year}{2022}{\natexlab{b}}), \eprint{2107.10285}.

\bibitem[{\citenamefont{Polchinski and Strassler}(2002)}]{Polchinski:2001tt}
\bibinfo{author}{\bibfnamefont{J.}~\bibnamefont{Polchinski}} \bibnamefont{and}
  \bibinfo{author}{\bibfnamefont{M.~J.} \bibnamefont{Strassler}},
  \bibinfo{journal}{Phys. Rev. Lett.} \textbf{\bibinfo{volume}{88}},
  \bibinfo{pages}{031601} (\bibinfo{year}{2002}), \eprint{hep-th/0109174}.

\bibitem[{\citenamefont{Bellazzini et~al.}(2020)\citenamefont{Bellazzini,
  Elias~Mir\'o, Rattazzi, Riembau, and Riva}}]{Bellazzini:2020cot}
\bibinfo{author}{\bibfnamefont{B.}~\bibnamefont{Bellazzini}},
  \bibinfo{author}{\bibfnamefont{J.}~\bibnamefont{Elias~Mir\'o}},
  \bibinfo{author}{\bibfnamefont{R.}~\bibnamefont{Rattazzi}},
  \bibinfo{author}{\bibfnamefont{M.}~\bibnamefont{Riembau}}, \bibnamefont{and}
  \bibinfo{author}{\bibfnamefont{F.}~\bibnamefont{Riva}}
  (\bibinfo{year}{2020}), \eprint{2011.00037}.

\bibitem[{\citenamefont{Tolley et~al.}(2021)\citenamefont{Tolley, Wang, and
  Zhou}}]{Tolley:2020gtv}
\bibinfo{author}{\bibfnamefont{A.~J.} \bibnamefont{Tolley}},
  \bibinfo{author}{\bibfnamefont{Z.-Y.} \bibnamefont{Wang}}, \bibnamefont{and}
  \bibinfo{author}{\bibfnamefont{S.-Y.} \bibnamefont{Zhou}},
  \bibinfo{journal}{JHEP} \textbf{\bibinfo{volume}{05}}, \bibinfo{pages}{255}
  (\bibinfo{year}{2021}), \eprint{2011.02400}.

\bibitem[{\citenamefont{Caron-Huot and Van~Duong}(2021)}]{Caron-Huot:2020cmc}
\bibinfo{author}{\bibfnamefont{S.}~\bibnamefont{Caron-Huot}} \bibnamefont{and}
  \bibinfo{author}{\bibfnamefont{V.}~\bibnamefont{Van~Duong}},
  \bibinfo{journal}{JHEP} \textbf{\bibinfo{volume}{05}}, \bibinfo{pages}{280}
  (\bibinfo{year}{2021}), \eprint{2011.02957}.

\bibitem[{\citenamefont{Sinha and Zahed}(2021)}]{Sinha:2020win}
\bibinfo{author}{\bibfnamefont{A.}~\bibnamefont{Sinha}} \bibnamefont{and}
  \bibinfo{author}{\bibfnamefont{A.}~\bibnamefont{Zahed}},
  \bibinfo{journal}{Phys. Rev. Lett.} \textbf{\bibinfo{volume}{126}},
  \bibinfo{pages}{181601} (\bibinfo{year}{2021}), \eprint{2012.04877}.

\bibitem[{\citenamefont{Arkani-Hamed
  et~al.}(2021{\natexlab{a}})\citenamefont{Arkani-Hamed, Huang, and
  Huang}}]{Arkani-Hamed:2020blm}
\bibinfo{author}{\bibfnamefont{N.}~\bibnamefont{Arkani-Hamed}},
  \bibinfo{author}{\bibfnamefont{T.-C.} \bibnamefont{Huang}}, \bibnamefont{and}
  \bibinfo{author}{\bibfnamefont{Y.-T.} \bibnamefont{Huang}},
  \bibinfo{journal}{JHEP} \textbf{\bibinfo{volume}{05}}, \bibinfo{pages}{259}
  (\bibinfo{year}{2021}{\natexlab{a}}), \eprint{2012.15849}.

\bibitem[{\citenamefont{Li et~al.}(2021)\citenamefont{Li, Xu, Yang, Zhang, and
  Zhou}}]{Li:2021lpe}
\bibinfo{author}{\bibfnamefont{X.}~\bibnamefont{Li}},
  \bibinfo{author}{\bibfnamefont{H.}~\bibnamefont{Xu}},
  \bibinfo{author}{\bibfnamefont{C.}~\bibnamefont{Yang}},
  \bibinfo{author}{\bibfnamefont{C.}~\bibnamefont{Zhang}}, \bibnamefont{and}
  \bibinfo{author}{\bibfnamefont{S.-Y.} \bibnamefont{Zhou}},
  \bibinfo{journal}{Phys. Rev. Lett.} \textbf{\bibinfo{volume}{127}},
  \bibinfo{pages}{121601} (\bibinfo{year}{2021}), \eprint{2101.01191}.

\bibitem[{\citenamefont{Caron-Huot
  et~al.}(2021{\natexlab{a}})\citenamefont{Caron-Huot, Mazac, Rastelli, and
  Simmons-Duffin}}]{Caron-Huot:2021rmr}
\bibinfo{author}{\bibfnamefont{S.}~\bibnamefont{Caron-Huot}},
  \bibinfo{author}{\bibfnamefont{D.}~\bibnamefont{Mazac}},
  \bibinfo{author}{\bibfnamefont{L.}~\bibnamefont{Rastelli}}, \bibnamefont{and}
  \bibinfo{author}{\bibfnamefont{D.}~\bibnamefont{Simmons-Duffin}},
  \bibinfo{journal}{JHEP} \textbf{\bibinfo{volume}{07}}, \bibinfo{pages}{110}
  (\bibinfo{year}{2021}{\natexlab{a}}), \eprint{2102.08951}.

\bibitem[{\citenamefont{Haldar et~al.}(2021)\citenamefont{Haldar, Sinha, and
  Zahed}}]{Haldar:2021rri}
\bibinfo{author}{\bibfnamefont{P.}~\bibnamefont{Haldar}},
  \bibinfo{author}{\bibfnamefont{A.}~\bibnamefont{Sinha}}, \bibnamefont{and}
  \bibinfo{author}{\bibfnamefont{A.}~\bibnamefont{Zahed}},
  \bibinfo{journal}{SciPost Phys.} \textbf{\bibinfo{volume}{11}},
  \bibinfo{pages}{002} (\bibinfo{year}{2021}), \eprint{2103.12108}.

\bibitem[{\citenamefont{Bern et~al.}(2021)\citenamefont{Bern, Kosmopoulos, and
  Zhiboedov}}]{Bern:2021ppb}
\bibinfo{author}{\bibfnamefont{Z.}~\bibnamefont{Bern}},
  \bibinfo{author}{\bibfnamefont{D.}~\bibnamefont{Kosmopoulos}},
  \bibnamefont{and}
  \bibinfo{author}{\bibfnamefont{A.}~\bibnamefont{Zhiboedov}},
  \bibinfo{journal}{J. Phys. A} \textbf{\bibinfo{volume}{54}},
  \bibinfo{pages}{344002} (\bibinfo{year}{2021}), \eprint{2103.12728}.

\bibitem[{\citenamefont{Chiang et~al.}(2021)\citenamefont{Chiang, Huang, Li,
  Rodina, and Weng}}]{Chiang:2021ziz}
\bibinfo{author}{\bibfnamefont{L.-Y.} \bibnamefont{Chiang}},
  \bibinfo{author}{\bibfnamefont{Y.-t.} \bibnamefont{Huang}},
  \bibinfo{author}{\bibfnamefont{W.}~\bibnamefont{Li}},
  \bibinfo{author}{\bibfnamefont{L.}~\bibnamefont{Rodina}}, \bibnamefont{and}
  \bibinfo{author}{\bibfnamefont{H.-C.} \bibnamefont{Weng}}
  (\bibinfo{year}{2021}), \eprint{2105.02862}.

\bibitem[{\citenamefont{Henriksson et~al.}(2021)\citenamefont{Henriksson,
  McPeak, Russo, and Vichi}}]{Henriksson:2021ymi}
\bibinfo{author}{\bibfnamefont{J.}~\bibnamefont{Henriksson}},
  \bibinfo{author}{\bibfnamefont{B.}~\bibnamefont{McPeak}},
  \bibinfo{author}{\bibfnamefont{F.}~\bibnamefont{Russo}}, \bibnamefont{and}
  \bibinfo{author}{\bibfnamefont{A.}~\bibnamefont{Vichi}}
  (\bibinfo{year}{2021}), \eprint{2107.13009}.

\bibitem[{\citenamefont{Chowdhury et~al.}(2021)\citenamefont{Chowdhury, Ghosh,
  Haldar, Raman, and Sinha}}]{Chowdhury:2021ynh}
\bibinfo{author}{\bibfnamefont{S.~D.} \bibnamefont{Chowdhury}},
  \bibinfo{author}{\bibfnamefont{K.}~\bibnamefont{Ghosh}},
  \bibinfo{author}{\bibfnamefont{P.}~\bibnamefont{Haldar}},
  \bibinfo{author}{\bibfnamefont{P.}~\bibnamefont{Raman}}, \bibnamefont{and}
  \bibinfo{author}{\bibfnamefont{A.}~\bibnamefont{Sinha}}
  (\bibinfo{year}{2021}), \eprint{2112.11755}.

\bibitem[{\citenamefont{Bellazzini et~al.}(2021)\citenamefont{Bellazzini,
  Riembau, and Riva}}]{Bellazzini:2021oaj}
\bibinfo{author}{\bibfnamefont{B.}~\bibnamefont{Bellazzini}},
  \bibinfo{author}{\bibfnamefont{M.}~\bibnamefont{Riembau}}, \bibnamefont{and}
  \bibinfo{author}{\bibfnamefont{F.}~\bibnamefont{Riva}}
  (\bibinfo{year}{2021}), \eprint{2112.12561}.

\bibitem[{\citenamefont{Caron-Huot et~al.}(2022)\citenamefont{Caron-Huot, Li,
  Parra-Martinez, and Simmons-Duffin}}]{Caron-Huot:2022ugt}
\bibinfo{author}{\bibfnamefont{S.}~\bibnamefont{Caron-Huot}},
  \bibinfo{author}{\bibfnamefont{Y.-Z.} \bibnamefont{Li}},
  \bibinfo{author}{\bibfnamefont{J.}~\bibnamefont{Parra-Martinez}},
  \bibnamefont{and}
  \bibinfo{author}{\bibfnamefont{D.}~\bibnamefont{Simmons-Duffin}}
  (\bibinfo{year}{2022}), \eprint{2201.06602}.

\bibitem[{\citenamefont{Caristo et~al.}(2021)\citenamefont{Caristo, Caselle,
  Magnoli, Nada, Panero, and Smecca}}]{Caristo:2021tbk}
\bibinfo{author}{\bibfnamefont{F.}~\bibnamefont{Caristo}},
  \bibinfo{author}{\bibfnamefont{M.}~\bibnamefont{Caselle}},
  \bibinfo{author}{\bibfnamefont{N.}~\bibnamefont{Magnoli}},
  \bibinfo{author}{\bibfnamefont{A.}~\bibnamefont{Nada}},
  \bibinfo{author}{\bibfnamefont{M.}~\bibnamefont{Panero}}, \bibnamefont{and}
  \bibinfo{author}{\bibfnamefont{A.}~\bibnamefont{Smecca}}
  (\bibinfo{year}{2021}), \eprint{2109.06212}.

\bibitem[{\citenamefont{Caron-Huot et~al.}(2017)\citenamefont{Caron-Huot,
  Komargodski, Sever, and Zhiboedov}}]{Caron-Huot:2016icg}
\bibinfo{author}{\bibfnamefont{S.}~\bibnamefont{Caron-Huot}},
  \bibinfo{author}{\bibfnamefont{Z.}~\bibnamefont{Komargodski}},
  \bibinfo{author}{\bibfnamefont{A.}~\bibnamefont{Sever}}, \bibnamefont{and}
  \bibinfo{author}{\bibfnamefont{A.}~\bibnamefont{Zhiboedov}},
  \bibinfo{journal}{JHEP} \textbf{\bibinfo{volume}{10}}, \bibinfo{pages}{026}
  (\bibinfo{year}{2017}), \eprint{1607.04253}.

\bibitem[{\citenamefont{Sever and Zhiboedov}(2018)}]{Sever:2017ylk}
\bibinfo{author}{\bibfnamefont{A.}~\bibnamefont{Sever}} \bibnamefont{and}
  \bibinfo{author}{\bibfnamefont{A.}~\bibnamefont{Zhiboedov}},
  \bibinfo{journal}{JHEP} \textbf{\bibinfo{volume}{06}}, \bibinfo{pages}{054}
  (\bibinfo{year}{2018}), \eprint{1707.05270}.

\bibitem[{\citenamefont{Chowdhury et~al.}(2020)\citenamefont{Chowdhury, Gadde,
  Gopalka, Halder, Janagal, and Minwalla}}]{Chowdhury:2019kaq}
\bibinfo{author}{\bibfnamefont{S.~D.} \bibnamefont{Chowdhury}},
  \bibinfo{author}{\bibfnamefont{A.}~\bibnamefont{Gadde}},
  \bibinfo{author}{\bibfnamefont{T.}~\bibnamefont{Gopalka}},
  \bibinfo{author}{\bibfnamefont{I.}~\bibnamefont{Halder}},
  \bibinfo{author}{\bibfnamefont{L.}~\bibnamefont{Janagal}}, \bibnamefont{and}
  \bibinfo{author}{\bibfnamefont{S.}~\bibnamefont{Minwalla}},
  \bibinfo{journal}{JHEP} \textbf{\bibinfo{volume}{02}}, \bibinfo{pages}{114}
  (\bibinfo{year}{2020}), \eprint{1910.14392}.

\bibitem[{\citenamefont{Huang et~al.}(2021)\citenamefont{Huang, Liu, Rodina,
  and Wang}}]{Huang:2020nqy}
\bibinfo{author}{\bibfnamefont{Y.-t.} \bibnamefont{Huang}},
  \bibinfo{author}{\bibfnamefont{J.-Y.} \bibnamefont{Liu}},
  \bibinfo{author}{\bibfnamefont{L.}~\bibnamefont{Rodina}}, \bibnamefont{and}
  \bibinfo{author}{\bibfnamefont{Y.}~\bibnamefont{Wang}},
  \bibinfo{journal}{JHEP} \textbf{\bibinfo{volume}{04}}, \bibinfo{pages}{195}
  (\bibinfo{year}{2021}), \eprint{2008.02293}.

\bibitem[{\citenamefont{Chandorkar et~al.}(2021)\citenamefont{Chandorkar,
  Chowdhury, Kundu, and Minwalla}}]{Chandorkar:2021viw}
\bibinfo{author}{\bibfnamefont{D.}~\bibnamefont{Chandorkar}},
  \bibinfo{author}{\bibfnamefont{S.~D.} \bibnamefont{Chowdhury}},
  \bibinfo{author}{\bibfnamefont{S.}~\bibnamefont{Kundu}}, \bibnamefont{and}
  \bibinfo{author}{\bibfnamefont{S.}~\bibnamefont{Minwalla}},
  \bibinfo{journal}{JHEP} \textbf{\bibinfo{volume}{05}}, \bibinfo{pages}{143}
  (\bibinfo{year}{2021}), \eprint{2102.03122}.

\bibitem[{\citenamefont{Mack}(2009)}]{Mack:2009mi}
\bibinfo{author}{\bibfnamefont{G.}~\bibnamefont{Mack}} (\bibinfo{year}{2009}),
  \eprint{0907.2407}.

\bibitem[{\citenamefont{Penedones}(2011)}]{Penedones:2010ue}
\bibinfo{author}{\bibfnamefont{J.}~\bibnamefont{Penedones}},
  \bibinfo{journal}{JHEP} \textbf{\bibinfo{volume}{03}}, \bibinfo{pages}{025}
  (\bibinfo{year}{2011}), \eprint{1011.1485}.

\bibitem[{\citenamefont{Paulos et~al.}(2017{\natexlab{b}})\citenamefont{Paulos,
  Penedones, Toledo, van Rees, and Vieira}}]{Paulos:2016fap}
\bibinfo{author}{\bibfnamefont{M.~F.} \bibnamefont{Paulos}},
  \bibinfo{author}{\bibfnamefont{J.}~\bibnamefont{Penedones}},
  \bibinfo{author}{\bibfnamefont{J.}~\bibnamefont{Toledo}},
  \bibinfo{author}{\bibfnamefont{B.~C.} \bibnamefont{van Rees}},
  \bibnamefont{and} \bibinfo{author}{\bibfnamefont{P.}~\bibnamefont{Vieira}},
  \bibinfo{journal}{JHEP} \textbf{\bibinfo{volume}{11}}, \bibinfo{pages}{133}
  (\bibinfo{year}{2017}{\natexlab{b}}), \eprint{1607.06109}.

\bibitem[{\citenamefont{Dubovsky et~al.}(2017)\citenamefont{Dubovsky, Gorbenko,
  and Mirbabayi}}]{Dubovsky:2017cnj}
\bibinfo{author}{\bibfnamefont{S.}~\bibnamefont{Dubovsky}},
  \bibinfo{author}{\bibfnamefont{V.}~\bibnamefont{Gorbenko}}, \bibnamefont{and}
  \bibinfo{author}{\bibfnamefont{M.}~\bibnamefont{Mirbabayi}},
  \bibinfo{journal}{JHEP} \textbf{\bibinfo{volume}{09}}, \bibinfo{pages}{136}
  (\bibinfo{year}{2017}), \eprint{1706.06604}.

\bibitem[{\citenamefont{Hijano}(2019)}]{Hijano:2019qmi}
\bibinfo{author}{\bibfnamefont{E.}~\bibnamefont{Hijano}},
  \bibinfo{journal}{JHEP} \textbf{\bibinfo{volume}{07}}, \bibinfo{pages}{132}
  (\bibinfo{year}{2019}), \eprint{1905.02729}.

\bibitem[{\citenamefont{Hijano and Neuenfeld}(2020)}]{Hijano:2020szl}
\bibinfo{author}{\bibfnamefont{E.}~\bibnamefont{Hijano}} \bibnamefont{and}
  \bibinfo{author}{\bibfnamefont{D.}~\bibnamefont{Neuenfeld}},
  \bibinfo{journal}{JHEP} \textbf{\bibinfo{volume}{11}}, \bibinfo{pages}{009}
  (\bibinfo{year}{2020}), \eprint{2005.03667}.

\bibitem[{\citenamefont{Komatsu et~al.}(2020)\citenamefont{Komatsu, Paulos,
  Van~Rees, and Zhao}}]{Komatsu:2020sag}
\bibinfo{author}{\bibfnamefont{S.}~\bibnamefont{Komatsu}},
  \bibinfo{author}{\bibfnamefont{M.~F.} \bibnamefont{Paulos}},
  \bibinfo{author}{\bibfnamefont{B.~C.} \bibnamefont{Van~Rees}},
  \bibnamefont{and} \bibinfo{author}{\bibfnamefont{X.}~\bibnamefont{Zhao}},
  \bibinfo{journal}{JHEP} \textbf{\bibinfo{volume}{11}}, \bibinfo{pages}{046}
  (\bibinfo{year}{2020}), \eprint{2007.13745}.

\bibitem[{\citenamefont{Mazac}(2017)}]{Mazac:2016qev}
\bibinfo{author}{\bibfnamefont{D.}~\bibnamefont{Mazac}},
  \bibinfo{journal}{JHEP} \textbf{\bibinfo{volume}{04}}, \bibinfo{pages}{146}
  (\bibinfo{year}{2017}), \eprint{1611.10060}.

\bibitem[{\citenamefont{Mazac and Paulos}(2019{\natexlab{a}})}]{Mazac:2018mdx}
\bibinfo{author}{\bibfnamefont{D.}~\bibnamefont{Mazac}} \bibnamefont{and}
  \bibinfo{author}{\bibfnamefont{M.~F.} \bibnamefont{Paulos}},
  \bibinfo{journal}{JHEP} \textbf{\bibinfo{volume}{02}}, \bibinfo{pages}{162}
  (\bibinfo{year}{2019}{\natexlab{a}}), \eprint{1803.10233}.

\bibitem[{\citenamefont{Mazac and Paulos}(2019{\natexlab{b}})}]{Mazac:2018ycv}
\bibinfo{author}{\bibfnamefont{D.}~\bibnamefont{Mazac}} \bibnamefont{and}
  \bibinfo{author}{\bibfnamefont{M.~F.} \bibnamefont{Paulos}},
  \bibinfo{journal}{JHEP} \textbf{\bibinfo{volume}{02}}, \bibinfo{pages}{163}
  (\bibinfo{year}{2019}{\natexlab{b}}), \eprint{1811.10646}.

\bibitem[{\citenamefont{Carmi et~al.}(2019)\citenamefont{Carmi, Di~Pietro, and
  Komatsu}}]{Carmi:2018qzm}
\bibinfo{author}{\bibfnamefont{D.}~\bibnamefont{Carmi}},
  \bibinfo{author}{\bibfnamefont{L.}~\bibnamefont{Di~Pietro}},
  \bibnamefont{and} \bibinfo{author}{\bibfnamefont{S.}~\bibnamefont{Komatsu}},
  \bibinfo{journal}{JHEP} \textbf{\bibinfo{volume}{01}}, \bibinfo{pages}{200}
  (\bibinfo{year}{2019}), \eprint{1810.04185}.

\bibitem[{\citenamefont{Giombi and Khanchandani}(2020)}]{Giombi:2020rmc}
\bibinfo{author}{\bibfnamefont{S.}~\bibnamefont{Giombi}} \bibnamefont{and}
  \bibinfo{author}{\bibfnamefont{H.}~\bibnamefont{Khanchandani}},
  \bibinfo{journal}{JHEP} \textbf{\bibinfo{volume}{11}}, \bibinfo{pages}{118}
  (\bibinfo{year}{2020}), \eprint{2007.04955}.

\bibitem[{\citenamefont{Kundu}(2020{\natexlab{a}})}]{Kundu:2019zsl}
\bibinfo{author}{\bibfnamefont{S.}~\bibnamefont{Kundu}},
  \bibinfo{journal}{JHEP} \textbf{\bibinfo{volume}{05}}, \bibinfo{pages}{014}
  (\bibinfo{year}{2020}{\natexlab{a}}), \eprint{1912.09479}.

\bibitem[{\citenamefont{Meltzer and Sivaramakrishnan}(2020)}]{Meltzer:2020qbr}
\bibinfo{author}{\bibfnamefont{D.}~\bibnamefont{Meltzer}} \bibnamefont{and}
  \bibinfo{author}{\bibfnamefont{A.}~\bibnamefont{Sivaramakrishnan}},
  \bibinfo{journal}{JHEP} \textbf{\bibinfo{volume}{11}}, \bibinfo{pages}{073}
  (\bibinfo{year}{2020}), \eprint{2008.11730}.

\bibitem[{\citenamefont{Antunes et~al.}(2021)\citenamefont{Antunes, Costa,
  Penedones, Salgarkar, and van Rees}}]{Antunes:2021abs}
\bibinfo{author}{\bibfnamefont{A.}~\bibnamefont{Antunes}},
  \bibinfo{author}{\bibfnamefont{M.~S.} \bibnamefont{Costa}},
  \bibinfo{author}{\bibfnamefont{J.}~\bibnamefont{Penedones}},
  \bibinfo{author}{\bibfnamefont{A.}~\bibnamefont{Salgarkar}},
  \bibnamefont{and} \bibinfo{author}{\bibfnamefont{B.~C.} \bibnamefont{van
  Rees}}, \bibinfo{journal}{JHEP} \textbf{\bibinfo{volume}{12}},
  \bibinfo{pages}{094} (\bibinfo{year}{2021}), \eprint{2109.13261}.

\bibitem[{\citenamefont{Haldar and Sinha}(2020)}]{Haldar:2019prg}
\bibinfo{author}{\bibfnamefont{P.}~\bibnamefont{Haldar}} \bibnamefont{and}
  \bibinfo{author}{\bibfnamefont{A.}~\bibnamefont{Sinha}},
  \bibinfo{journal}{SciPost Phys.} \textbf{\bibinfo{volume}{8}},
  \bibinfo{pages}{095} (\bibinfo{year}{2020}), \eprint{1911.05974}.

\bibitem[{\citenamefont{Kundu}(2020{\natexlab{b}})}]{Kundu:2020bdn}
\bibinfo{author}{\bibfnamefont{S.}~\bibnamefont{Kundu}}
  (\bibinfo{year}{2020}{\natexlab{b}}), \eprint{2012.10450}.

\bibitem[{\citenamefont{Caron-Huot
  et~al.}(2021{\natexlab{b}})\citenamefont{Caron-Huot, Mazac, Rastelli, and
  Simmons-Duffin}}]{Caron-Huot:2021enk}
\bibinfo{author}{\bibfnamefont{S.}~\bibnamefont{Caron-Huot}},
  \bibinfo{author}{\bibfnamefont{D.}~\bibnamefont{Mazac}},
  \bibinfo{author}{\bibfnamefont{L.}~\bibnamefont{Rastelli}}, \bibnamefont{and}
  \bibinfo{author}{\bibfnamefont{D.}~\bibnamefont{Simmons-Duffin}},
  \bibinfo{journal}{JHEP} \textbf{\bibinfo{volume}{11}}, \bibinfo{pages}{164}
  (\bibinfo{year}{2021}{\natexlab{b}}), \eprint{2106.10274}.

\bibitem[{\citenamefont{Caron-Huot
  et~al.}(2021{\natexlab{c}})\citenamefont{Caron-Huot, Mazac, Rastelli, and
  Simmons-Duffin}}]{Caron-Huot:2020adz}
\bibinfo{author}{\bibfnamefont{S.}~\bibnamefont{Caron-Huot}},
  \bibinfo{author}{\bibfnamefont{D.}~\bibnamefont{Mazac}},
  \bibinfo{author}{\bibfnamefont{L.}~\bibnamefont{Rastelli}}, \bibnamefont{and}
  \bibinfo{author}{\bibfnamefont{D.}~\bibnamefont{Simmons-Duffin}},
  \bibinfo{journal}{JHEP} \textbf{\bibinfo{volume}{05}}, \bibinfo{pages}{243}
  (\bibinfo{year}{2021}{\natexlab{c}}), \eprint{2008.04931}.

\bibitem[{\citenamefont{Wilczek}(1982)}]{Wilczek:1982wy}
\bibinfo{author}{\bibfnamefont{F.}~\bibnamefont{Wilczek}},
  \bibinfo{journal}{Phys. Rev. Lett.} \textbf{\bibinfo{volume}{49}},
  \bibinfo{pages}{957} (\bibinfo{year}{1982}).

\bibitem[{\citenamefont{Jain et~al.}(2015)\citenamefont{Jain, Mandlik,
  Minwalla, Takimi, Wadia, and Yokoyama}}]{Jain:2014nza}
\bibinfo{author}{\bibfnamefont{S.}~\bibnamefont{Jain}},
  \bibinfo{author}{\bibfnamefont{M.}~\bibnamefont{Mandlik}},
  \bibinfo{author}{\bibfnamefont{S.}~\bibnamefont{Minwalla}},
  \bibinfo{author}{\bibfnamefont{T.}~\bibnamefont{Takimi}},
  \bibinfo{author}{\bibfnamefont{S.~R.} \bibnamefont{Wadia}}, \bibnamefont{and}
  \bibinfo{author}{\bibfnamefont{S.}~\bibnamefont{Yokoyama}},
  \bibinfo{journal}{JHEP} \textbf{\bibinfo{volume}{04}}, \bibinfo{pages}{129}
  (\bibinfo{year}{2015}), \eprint{1404.6373}.

\bibitem[{\citenamefont{Kulish and Faddeev}(1970)}]{Kulish:1970ut}
\bibinfo{author}{\bibfnamefont{P.~P.} \bibnamefont{Kulish}} \bibnamefont{and}
  \bibinfo{author}{\bibfnamefont{L.~D.} \bibnamefont{Faddeev}},
  \bibinfo{journal}{Theor. Math. Phys.} \textbf{\bibinfo{volume}{4}},
  \bibinfo{pages}{745} (\bibinfo{year}{1970}).

\bibitem[{\citenamefont{Arkani-Hamed
  et~al.}(2021{\natexlab{b}})\citenamefont{Arkani-Hamed, Pate, Raclariu, and
  Strominger}}]{Arkani-Hamed:2020gyp}
\bibinfo{author}{\bibfnamefont{N.}~\bibnamefont{Arkani-Hamed}},
  \bibinfo{author}{\bibfnamefont{M.}~\bibnamefont{Pate}},
  \bibinfo{author}{\bibfnamefont{A.-M.} \bibnamefont{Raclariu}},
  \bibnamefont{and}
  \bibinfo{author}{\bibfnamefont{A.}~\bibnamefont{Strominger}},
  \bibinfo{journal}{JHEP} \textbf{\bibinfo{volume}{08}}, \bibinfo{pages}{062}
  (\bibinfo{year}{2021}{\natexlab{b}}), \eprint{2012.04208}.

\end{thebibliography}

\end{document}